\documentstyle[graphics,graphicx,times]{mn2e}
\input{psfig}

\begin{document}

\title[The Proximity Effect at $z>4$]{
Evidence for overdensity around $z_{\rm em}$~$>$~4 quasars from the 
proximity effect\thanks{Based on observations carried out at the Keck 
Telescope}
}

\def\inst#1{{${}^{#1}$}}

\author[Guimar\~aes et al.]{R.~Guimar\~aes\inst{1,2}, P.~Petitjean \inst{2,3}, 
E.~Rollinde\inst{2}, R. R. de Carvalho\inst{4}, S. G. Djorgovski\inst{5},  
\newauthor R. Srianand\inst{6}, A. Aghaee\inst{2,7}, and S. Castro\inst{5,8}  \\
$^1$ Observatorio Nacional - MCT, R. Gal. Jos\'e
    Cristino, 77, 20921-400, Rio de Janeiro, RJ - Brasil\\
$^2$ Institut d'Astrophysique de Paris \& Universit\'e Pierre et Marie Curie, 98 bis boulevard
        d'Arago, 75014 Paris, France \\
 $^3$ LERMA, Observatoire de Paris, 61 avenue de l'Observatoire, 75014, Paris,France \\
 $^4$ Instituto Nacional de Pesquisas Espaciais - 
INPE, Av. dos Astronautas, 1758, 12227-010, S. J. dos Campos, SP - Brasil\\
 $^5$ Palomar Observatory, California Institute of Technology, 105-24, 
Pasadena, CA 91125, USA\\
 $^6$ Inter University Center for Astronomy and Astrophysics, IUCAA, Post bag 4, Ganeshkhind, 
Pune 411 007, India\\
 $^7$ Department of Physics, University of Sistan and Baluchestan, 98135 Zahedan, Iran\\
 $^8$ European Southern Observatory, Karl-Schwarzschild Strasse, 2, Garching, Germany}
\date{Typeset \today ; Received / Accepted}
\maketitle

\begin{abstract}
We study the density field around $z_{\rm em}$~$>$~4 quasars using high quality medium spectral 
resolution ESI-Keck spectra ($R$~$\sim$~4300, SNR~$>$~25) of 45 high-redshift quasars selected from 
a total of 95 spectra. This large sample considerably increases the statistics compared to 
previous studies. The redshift evolution of the mean photo-ionization rate and the median optical 
depth of the intergalactic medium (IGM) are derived statistically from the observed transmitted 
flux and the pixel optical depth probability distribution function respectively. This is used 
to study the so-called proximity effect, that is, the observed decrease of the median optical 
depth of the IGM in the vicinity of the quasar caused by enhanced photo-ionization rate due 
to photons emitted by the quasar. We show that the proximity effect is correlated with the luminosity 
of the quasars, as expected. By comparing the observed decrease of the median optical depth with 
the theoretical expectation we find that the optical depth does not decrease as rapidly as 
expected when approaching the quasar if the gas in its vicinity is part of the standard IGM. 
We interpret this effect as revealing gaseous overdensities on scales as large as 
$\sim$15$h^{-1}$~Mpc. The mean overdensity is of the order of two and five within, respectively, 
10 and 3$h^{-1}$~Mpc. If true, this would indicate that high redshift quasars are located 
in the center of overdense regions that could evolve with time into  massive clusters of galaxies. 
The overdensity is correlated with luminosity: brighter quasars show higher overdensities.

\end{abstract}


\section{Introduction}

\hyphenation{author another created financial paper re-commend-ed Post-Script}

The Inter-Galactic Medium (IGM) has been intensively studied
using the absorption seen in the spectra of
quasi-stellar objects (QSOs) over a large redshift range ($0.16 \le
z_{\rm em} \le 6.3$).
This absorption, first identified by Lynds (1971), breaks up 
at high spectral resolution in hundreds of discrete absorption lines 
from, predominantly, HI Lyman  
UV resonance lines redshifted in a expanding universe (the so-called 
Ly$\alpha$ forest, see Rauch 1998 for a review). 

The Ly$\alpha$ forest was interpreted by Sargent et al. (1980) as the signature of
intervening HI clouds of cosmological nature embedded in a 
diffuse hot medium. The clouds were further 
described by Rees (1986) as gravitationally confined by dark-matter 
mini-halos.
The advent of numerical simulations has introduced a new and more general
scheme in which the IGM is a crucial element
of large scale structures and galaxy formation. 
It is now believed that the space distribution of
the gas traces the potential wells of the dark matter. In addition, most
of the baryons are in the IGM at high redshift, making the 
IGM the reservoir of gas for galaxy formation.
The numerical $N$-body simulations have been successful at reproducing the observed characteristics 
of the Ly$\alpha$ forest
(e.g., Cen et al. 1994; Petitjean et al. 1995; Hernquist et al. 1996; Theuns et al. 1998).
The IGM is therefore seen as a smooth pervasive medium which can be used to 
study the spatial distribution of the mass on scales larger than the Jeans' length.
This idea is reinforced by observations of multiple lines of sight
(e.g., Coppolani et al. 2006).

It is well known that the characteristics of the Ly$\alpha$ forest
change in the vicinity of the quasar due to the additional ionizing flux
produced by the quasar. The mean neutral hydrogen fraction decreases when
approaching the quasar. Because the 
amount of absorption in the IGM is, in general, increasing with redshift, this reversal in the 
cosmological trend for redshifts close to the emission redshift of the quasar is called the 'inverse' 
or 'proximity' effect (Carswell et al. 1982; Murdoch et al. 1986). 
It is possible to use this effect, together with a knowledge of the quasar
luminosity and its position, to derive the mean flux of the UV background
if one assumes that the redshift evolution of the density field can be 
extrapolated from far away to close to the quasar.
Indeed, the strength of the effect depends on the ratio of the ionization rates 
from the quasar and the UV background, and because the quasar's ionization 
rate can be determined directly through the knowledge of its luminosity
and distance, the ionization rate in the IGM can be inferred. This method was 
pioneered by Bajtlik, Duncan \& Ostriker (1988) but more 
recent data have yielded a wide variety of estimates (Lu, Wolfe \& Turnshek 1991; 
Kulkarni \& Fall 1993; Bechtold 1994; Cristiani et al. 1995; 
Fernandez-Soto et al. 1995; Giallongo et al. 1996;  
Srianand \& Khare 1996; Cooke, Espey \& Carswell 1997; Scott et al. 2000, 2002; 
Liske \& Williger 2001). Scott et al. (2000) collected estimates from the 
literature which vary over almost an order of magnitude at $z= 3$. 
The large scatter in the results can be explained by errors in the continuum placement, 
cosmic variance, redshift determination, etc. 

In the standard analysis of the proximity effect it is assumed that the underlying 
matter distribution is not altered by the presence of the quasar.
The only difference between the gas close to the quasar or far away from it is the increased 
photoionization rate in the vicinity of the QSO. 
If true, the strength of 
the proximity effect should correlate with the quasar luminosity but such a 
correlation has not been convincingly established (see Lu et al. 1991; Bechtold 1994; 
Srianand \& Khare 1996; see however Liske \& Williger 2001).
It is in fact likely that 
the quasars are located inside overdense regions. Indeed, the presence of Ly$\alpha$
absorption lines with $z_{\rm abs}$~$>$~$z_{\rm em}$ 
suggests an excess of material around QSOs (Loeb \& Eisenstein 1995; 
Srianand \& Khare 1996). Furthermore, in hierarchical models of galaxy formation, 
the supermassive black holes that are thought to power quasars are located in massive 
haloes (Magorrian et al. 1998; Ferrarese 2002), that are strongly 
biased to high-density regions. 
Possible evidence for overdensities around quasars come also from 
studies of the transverse proximity effect by Croft (2004), Schirber, 
Miralda-Escud\'e \& McDonald (2004) and Worseck \& Wisotzki (2006) who suggest 
that the observed absorption is larger than that predicted by models assuming standard 
proximity effect and isotropic quasar emission. 
However, in the case of transverse observations, it could be that the quasar light is 
strongly beamed in our direction or, alternatively, that the quasar is highly variable. 
Interestingly, neither of these affects the longitudinal proximity effect discussed in the
present paper.

Observations of the IGM transmission close to Lyman break galaxies (LBGs) seem to show that,
close to the galaxy, the IGM contains more neutral hydrogen than on average
(Adelberger et al. 2003). As the UV photons from the LBGs cannot alter the ionization 
state of the gas at large distances, it is most likely that the excess absorption 
is caused by the enhanced IGM density around LBGs. It is worth noting however that various 
hydrodynamical simulations have trouble reproducing this so-called galaxy proximity 
effect (e.g., Bruscoli et al. 2003; Kollmeier et al. 2003; Maselli et al. 2004;
Desjacques et al. 2004). 

In a recent paper, Rollinde et al. (2005) presented a new analysis to infer the 
density structure around quasars. The method is based on the determination of the
cumulative probability distribution function (CPDF) of pixel optical depth,
and so avoids the Voigt profile fitting and line counting which is traditionally 
used (e.g., Cowie \& Songaila 1998; Ellison et al. 2000; Aguirre et al. 2002; 
Schaye et al 2003; Aracil et al. 2004; Pieri et al. 2006).  
The evolution in redshift of the optical depth CPDF far away from the quasar 
is directly derived from the data. This redshift dependent
CPDF is then compared to the CPDF observed close to the quasar to
derive the mean density profile around quasars. The method was
applied to twenty lines of sight toward quasars at 
$z_{\rm em}\sim 2$ observed with UVES/VLT and it was found that overdensities 
of the order of a few are 
needed for the observations to be consistent with the value of the UV background 
flux derived from the mean Ly$\alpha$ opacity. In the present paper we apply the same 
method to a large sample of 95 quasars at $z_{\rm em}$~$>$~4 observed with the Echelle 
Spectrograph and Imager (ESI) mounted on the Keck II telescope.

In Section~2 we describe the data and the selection of the sample used
in the present work. We derive the redshift evolution of the ionizing
UV background and of the median IGM optical depth in Sections 3 and 4
respectively. We discuss the proximity effect in Section~5
and conclude in Section~6.
We assume throughout this paper a flat Universe with $\Omega_{\rm m}$~=~0.3,
$\Omega_{\rm \Lambda}$~=~0.7, $\Omega_{\rm b}$~=~0.04 and H$_o$~=~70~kms$^{-1}$.

\section{Data and sample selection}
Medium resolution (R~$\sim$~4300) 
spectra of all $z>3$ quasars discovered in the course of the DPOSS
survey (Digital Palomar Observatory Sky Survey; see, e.g., Kennefick et al. 1995, Djorgovski et al. 1999 and the complete listing of QSOs available at
http://www.astro.caltech.edu/$\sim$george/z4.qsos) have been 
obtained with the Echellette Spectrograph and Imager (ESI, Sheinis et al 2002) mounted on 
the KECK II 10~m telescope. Signal-to-noise ratio is usually larger than 15 per 10~km~s$^{-1}$ 
pixel .
These data have already been used to construct a sample of Damped Ly$\alpha$
systems at high redshift (Prochaska et al. 2003a,b). In total, 95 quasars have
been observed.

The KECK/ESI spectra were reduced (bias subtraction, flat-fielding, spectrum 
extraction) using standard procedures of the IRAF package. The different orders of 
the spectra were combined using the scombine task. In the Echellette mode spectra are 
divided in ten orders  covering the 
wavelength range: $4000$~\AA~$ \le \lambda_{\rm obs} \le$~10000~\AA.
In this work we use only orders 3 (center 4650~\AA) to 7 (center 6750~\AA). 
The spectral resolution is $R\sim4300$ or $\sim70$~km~s$^{-1}$. 
During the process of combining the orders, we controlled carefully 
the signal-to-noise ratio obtained in each order.
Wavelengths and redshifts were computed in the heliocentric 
restframe and the spectra were flux calibrated and then normalized.

The signal-to-noise ratio per pixel was obtained in the regions of
the Ly-$\alpha$ forest that are free of absorption and the mean SNR value, 
averaged between the Ly-$\alpha$ and Ly-$\beta$ emission lines, was computed. 
We used only spectra with mean SNR~$\ge$~25.
We rejected the broad absorption line QSOs (BAL) and
QSOs with more than one damped Ly$\alpha$ system (DLA) redshifted
between the Ly-$\alpha$ and the Ly-$\beta$ emission lines 
because their presence may over-pollute the 
Ly$\alpha$ forest. Metal absorptions are not subtracted from 
the spectra. We can estimate that the number of intervening CIV
and MgII systems with $W_{\rm obs}$~$>$~0.25~\AA~ is of the order
of five along each of the lines of sight (see e.g., Boksenberg et al. 2003; 
Aracil et al. 2004; Tytler et al. 2004; Scannapieco et al. 2006). This means we expect
an error on the determination of the mean absorption of the order of
1~\%. This is at least five times smaller than the error expected from
the placement of the continuum. 

\begin{figure}
\begin{center}
\includegraphics[height=8.5cm,width=8.5cm]{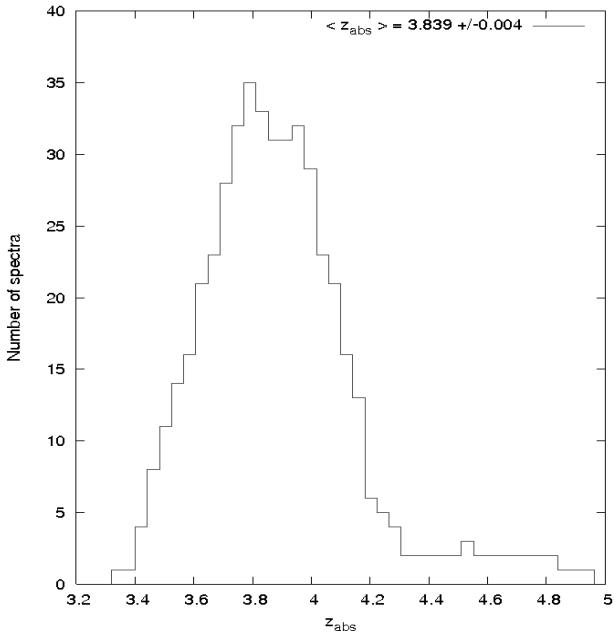}
\caption{Number of spectra in our sample contributing to the 
study of the Ly-$\alpha$ forest at a given redshift as a function of
redshift.}\label{Nxzem}
\end{center}
\end{figure}

\begin{figure*}
\begin{center}
\includegraphics[height=3.5cm,width=16cm]{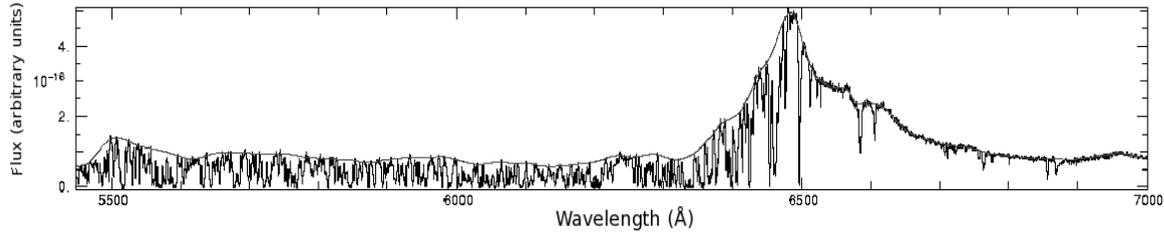}
\caption{Flux calibrated spectrum of PSS~1058+1245. The continuum
derived automatically is overplotted.
}\label{Spectra}
\end{center}
\end{figure*}

In Table \ref{logbook} we give the characteristics of the forty-five QSO
spectra satisfying the above criteria that are used in the present work. 
Column (1) gives the QSO's name; column (2) the emission redshift 
estimated as the average 
of the determinations of the peak of the Ly$\alpha$ emission and the peak
of a Gaussian fitted to the CIV emission line; 
column (3) the emission redshift obtained using the IRAF task rvidlines
(see Section 5.1);
column (4) the apparent V magnitude; 
column (5) the mean signal-to-noise ratio in the Ly$\alpha$ forest and 
column (6) the intrinsic luminosity
at the Lyman limit estimated from the V-magnitude assuming that the QSO continuum
spectrum is a power-law of index $-0.6$ (see Section~5).
In Fig.~\ref{Nxzem} we show the histogram of the number of spectra in our 
sample contributing to the study of the Ly$\alpha$ forest at a given 
redshift as a function of redshift.
Normalisation of the spectra is known to be a crucial step in these
studies. An automatic procedure  (Aracil et al. 2004) estimates iteratively  
the  continuum  by minimising the sum  of a regularisation term 
(the effect of which is to smooth the continuum)
and a $\chi^2$ term, which is computed from the difference between the 
quasar spectrum and the continuum estimated during the previous iteration.
Absorption lines are avoided when computing the continuum.
A few obvious defects are then corrected by hand adjusting the 
reference points of the fit. This happens to be important 
near the peak of strong emission lines and over damped absorption lines. 
The automatic method works very well and a minimal manual intervention is necessary.
The procedure was calibrated by Aracil et al. (2004) using simulated quasar spectra
(with emission and absorption lines) adding  continuum
modulations to  mimic an imperfect  correction of the blaze  along the
orders  and  noise to  obtain  a  S/N ratio  similar to  that in  the
data. We noted that the procedure underestimates  
the true continuum in the Ly$\alpha$ forest by a quantity depending 
smoothly on the wavelength and
the emission redshift by an amount of about 3-5\%  at $z$~$\sim$~3.5-4. 
(see Fig.~1 of Aracil et al. 2004). This is less than our typical errors
and therefore we did not correct the normalized spectra for this.
Note that, due to strong blending, errors can be as high as 10\% in places (see also Croft et al 2002; Becker et al. 2006; McDonald et al. 2006; Desjacques, Nusser \& Sheth 2007). 
In that case however, the optical depth will be usually larger than 
the limit we will used to reject pixels badly affected by saturation. 
It must be noted also that, because of the reduced absorption in the
vicinity of the quasar, the continuum determination is more reliable 
in the wing of the Ly$\alpha$ emission line.
The normalization procedure we have used is preferred to other 
methods that are more arbitrary. It is completely reproducible and introduces 
only a small systematic error that we can control.
A final sample reduced spectrum with the continuum fitting is shown in Fig.~\ref{Spectra}.

\begin{table*}
\begin{center}
\footnotesize \small
\caption{Quasar sample}
\label{logbook} \vspace{1pc}
\begin{tabular}{l l l l l c}
\hline
QSO & $z_{\rm em}$~$^a$ & $z_{\rm em}$~$^b$ & $V_{\rm mag}$ &  SNR & $L_{912}$~$^c$  \\
  &  &  & & & $h^{-2}$~ergs~s$^{-1}$Hz$^{-1}$\\
\hline
PSS0117+1552  &  4.241 & 4.244 & 18.6   & 45 & 4.53$\times $ 10$^{31}$\\
PSS0118+0320  &  4.235 & 4.232 & 18.50  & 30 & 4.38$\times $ 10$^{31}$\\
PSS0121+0347  &  4.130 & 4.127 & 17.86  & 72 & 6.76$\times $ 10$^{31}$\\
SDSS0127-0045 &  4.067 & 4.084 & 18.37  & 25 & 4.04$\times $ 10$^{31}$\\
PSS0131+0633  &  4.432 & 4.430 & 18.24  & 25 & 6.88$\times $ 10$^{31}$\\
PSS0134+3307  &  4.534 & 4.536 & 18.82  & 30 & 4.73$\times $ 10$^{31}$\\
PSS0209+0517  &  4.206 & 4.194 & 17.36  & 40 & 8.16$\times $ 10$^{31}$\\
PSS0211+1107  &  3.975 & 3.975 & 18.12  & 66 & 6.14$\times $ 10$^{31}$\\
PSS0248+1802  &  4.427 & 4.430 & 18.4   & 60 & 8.32$\times $ 10$^{31}$\\
PSS0452+0355  &  4.397 & 4.395 & 18.80  & 30 & 4.77$\times $ 10$^{31}$\\
PSS0747+4434  &  4.434 & 4.435 & 18.06  & 54 & 7.98$\times $ 10$^{31}$\\
PSS0808+5215  &  4.476 & 4.510 & 18.82  & 37 & 4.29$\times $ 10$^{31}$\\
SDSS0810+4603 &  4.078 & 4.074 & 18.67  & 43 & 3.22$\times $ 10$^{31}$\\
PSS0852+5045  &  4.213 & 4.216 & 19     & 54 & 2.60$\times $ 10$^{31}$\\
PSS0926+3055  &  4.188 & 4.198 & 17.31  & 77 & 1.22$\times $ 10$^{32}$\\
PSS0950+5801  &  3.969 & 3.973 & 17.38  & 85 & 9.31$\times $ 10$^{31}$\\
PSS0957+3308  &  4.283 & 4.274 & 17.59  & 40 & 9.80$\times $ 10$^{31}$\\
PSS1057+4555  &  4.127 & 4.126 & 17.7   & 76 & 7.67$\times $ 10$^{31}$\\
PSS1058+1245  &  4.332 & 4.330 & 18     & 46 & 7.45$\times $ 10$^{31}$\\
PSS1140+6205  &  4.507 & 4.509 & 18.73  & 61 & 4.77$\times $ 10$^{31}$\\
PSS1159+1337  &  4.089 & 4.081 & 18.50  & 45 & 3.65$\times $ 10$^{31}$\\
PSS1248+3110  &  4.358 & 4.346 & 18.9   & 25 & 3.19$\times $ 10$^{31}$\\
SDSS1310-0055 &  4.152 & 4.151 & 18.85  & 38 & 2.82$\times $ 10$^{31}$\\
PSS1317+3531  &  4.370 & 4.369 & 19.10  & 28 & 2.75$\times $ 10$^{31}$\\
J1325+1123    &  4.408 & 4.400 & 18.77  & 25 & 4.20$\times $ 10$^{31}$\\
PSS1326+0743  &  4.121 & 4.123 & 17.3   & 66 & 1.01$\times $ 10$^{32}$\\
PSS1347+4956  &  4.597 & 4.560 & 17.9   & 40 & 1.02$\times $ 10$^{32}$\\
PSS1401+4111  &  4.008 & 4.026 & 18.62  & 30 & 2.90$\times $ 10$^{31}$\\
PSS1403+4126  &  3.862 & 3.862 & 18.92  & 25 & 1.85$\times $ 10$^{31}$\\
PSS1430+2828  &  4.309 & 4.306 & 19.30  & 40 & 2.18$\times $ 10$^{31}$\\
PSS1432+3940  &  4.291 & 4.292 & 18.6   & 36 & 4.04$\times $ 10$^{31}$\\
PSS1443+2724  &  4.419 & 4.406 & 19.3   & 30 & 2.57$\times $ 10$^{31}$\\
PSS1458+6813  &  4.295 & 4.291 & 18.67  & 60 & 5.04$\times $ 10$^{31}$\\
PSS1500+5829  &  4.229 & 4.224 & 18.6   & 40 & 3.74$\times $ 10$^{31}$\\
GB1508+5714   &  4.306 & 4.304 & 18.9   & 32 & 3.12$\times $ 10$^{31}$\\
PSS1535+2943  &  3.979 & 3.972 & 18.9   & 29 & 2.26$\times $ 10$^{31}$\\
PSS1555+2003  &  4.226 & 4.228 & 18.9   & 31 & 3.15$\times $ 10$^{31}$\\
PSS1633+1411  &  4.360 & 4.349 & 19.0   & 43 & 3.36$\times $ 10$^{31}$\\
PSS1646+5514  &  4.110 & 4.084 & 18.11  & 37 & 4.81$\times $ 10$^{31}$\\
PSS1721+3256  &  4.031 & 4.040 & 19.23  & 43 & 1.85$\times $ 10$^{31}$\\
PSS1723+2243  &  4.515 & 4.514 & 18.17  & 42 & 8.77$\times $ 10$^{31}$\\
PSS2154+0335  &  4.349 & 4.359 & 18.41  & 26 & 6.20$\times $ 10$^{31}$\\
PSS2203+1824  &  4.372 & 4.375 & 18.74  & 34 & 4.50$\times $ 10$^{31}$\\
PSS2238+2603  &  4.023 & 4.031 & 18.85  & 24 & 2.74$\times $ 10$^{31}$\\
PSS2344+0342  &  4.341 & 4.340 & 17.87  & 30 & 4.38$\times $ 10$^{31}$\\
\hline\\
\multicolumn{6}{l}{$^a$ Mean of estimates from a Gaussian fit to the CIV
emission line and from}\\
\multicolumn{6}{l}{the peak of the Ly$\alpha$ emission}\\
\multicolumn{6}{l}{$^b$ Estimate using the IRAF RVIDLINES task}\\
\multicolumn{6}{l}{$^c$ continuum luminosity at 912~\AA}\\
\end{tabular}
\end{center}
\end{table*}

\section{The Ionizing Background radiation field}

The standard treatment of the proximity effect consists in studying the
evolution of the mean absorption in the Ly$\alpha$ forest
when approaching the quasar. Far away from the quasar the only source of 
ionizing photons is the UV background whereas in the vicinity of the quasar, 
the gas is ionized by both the UV background and the QSO. Assuming (i) that the 
luminosity of the quasar is known, (ii) that the distance from the gas to the quasar 
is cosmological (and therefore given by the difference in redshift)
and (iii) that the density field in the IGM is not modified by the presence
of the quasar, it is possible to derive observationally the distance to the quasar where 
the ionizing flux from the quasar equals the flux from the UV background.
This, in turn, gives an estimate of the UV background flux.
The last assumption neglects the fact that quasars can be surrounded by significant 
overdensities (e.g., Pascarelle et al. 2001; Adelberger et al. 2003; Rollinde et al. 2005; Faucher-Gigu\`ere et al. 2006; Kim \& Croft 2006).

However the above approach can be reversed to derive the density distribution
around the quasar if the UV background can be estimated from elsewhere.
Actually, it is possible to estimate this flux by modelling the redshift evolution of
the mean absorption in the IGM. 
For $z>4$ we can follow Songaila \& Cowie (2002) and use the mean normalized transmitted
flux, $F(z)$, to derive the normalized ionization rate, $g$, in units of 10$^{-12}$~s$^{-1}$,
by inverting the following equation (see also McDonald \& Miralda-Escud\'e 2001; 
Cen \& McDonald 2002):
\begin{equation}
\label{gammabg} F(z) = 4.5 \times g^{-0.28} \times
[\frac{(1+z)}{7}]^{2.2} \times e^{[-4.4 g^{-0.4} [\frac{(1+z)}{7}]^3]}
\end{equation}
with
\begin{equation}
\label{geq} g\equiv \Gamma_{-12} \times T_4^{0.75} \times h \times
(\frac{\Omega_{\rm m}}{0.35})^{0.5} \times (\frac{\Omega_{\rm b} h^2}{0.0325})^{-2}
\end{equation}

\begin{figure}
\begin{center}
\includegraphics[height=8.5cm,width=8.5cm]{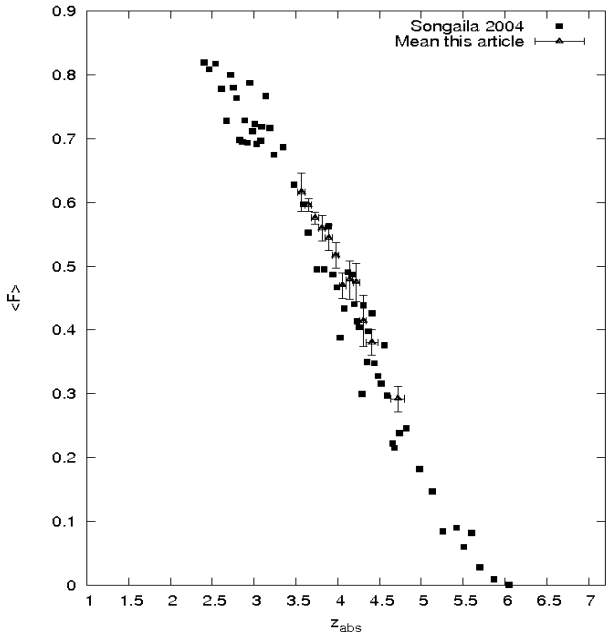}
\caption{The observed normalized transmitted flux $<F>$ as a function of redshift.}\label{meanflux}
\end{center}
\end{figure}

Applying the above equations, we can derive the redshift dependence of 
$\Gamma_{-12}$ from the observed transmitted flux (for $z>4$). For this, we have used the 
Ly$\alpha$ forest over the rest-wavelength range 1070-1170~\AA~ to avoid contamination
by the proximity effect close to the Ly$\alpha$ emission line
and by possible OVI associated absorbers. We also carefully avoided regions flagged
because of data reduction problems or damped Ly$\alpha$ absorption lines. 
We divided each spectrum in bins of length $50$~\AA~ in the observed frame, 
corresponding to about $\triangle z = 0.04$. The mean transmitted flux was calculated as the 
mean flux over all pixels in a bin. At each redshift, we then averaged the transmitted 
fluxes over all spectra covering this redshift.
Errors were estimated as the standard deviation of the mean values divided by 
the square root of the number of spectra. The corresponding scatter in the transmitted 
flux cannot be explained by photon noise or by uncertainties in the continuum, suggesting that 
errors are dominated by cosmic variance. The mean transmitted flux that we have obtained 
is plotted in Fig~\ref{meanflux} together with results by Songaila (2004).

\begin{figure}
\begin{center}
\includegraphics[height=8.5cm,width=8.5cm]{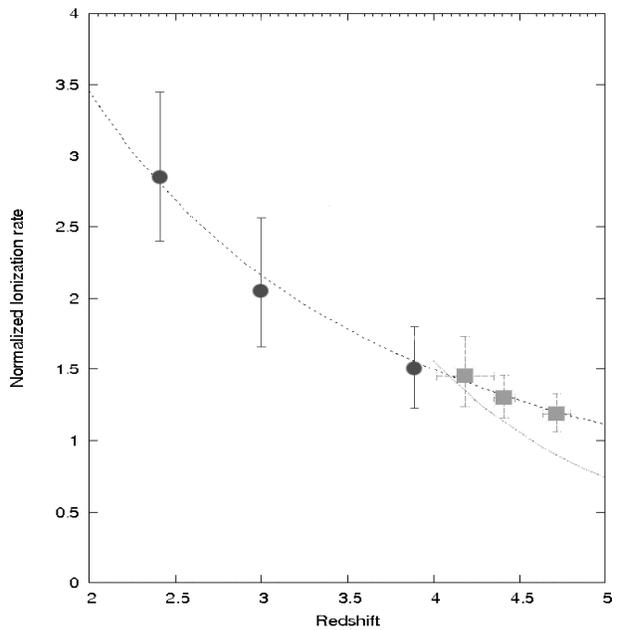}
\caption{The normalized ionization rate obtained from the 
transmitted flux observed in our data
by inverting Eq.~\ref{gammabg} in three
redshift bins (filled squares) at $z>4$. Filled circles are the measurements of
McDonald et al. (2000) for lower redshift. The dotted line is a
power law of the form $1.11\times[(1+z)/6]^{-1.63}$ fitted to the data
of McDonald et al. (2000) and this paper. The dashed line is a
power law of the form $0.74\times[(1+z)/6]^{-4.1}$ obtained, for $z>4$, by
Songaila (2004).}\label{g}
\end{center}
\end{figure}

\begin{figure}
\begin{center}
\includegraphics[height=8.5cm,width=8.5cm]{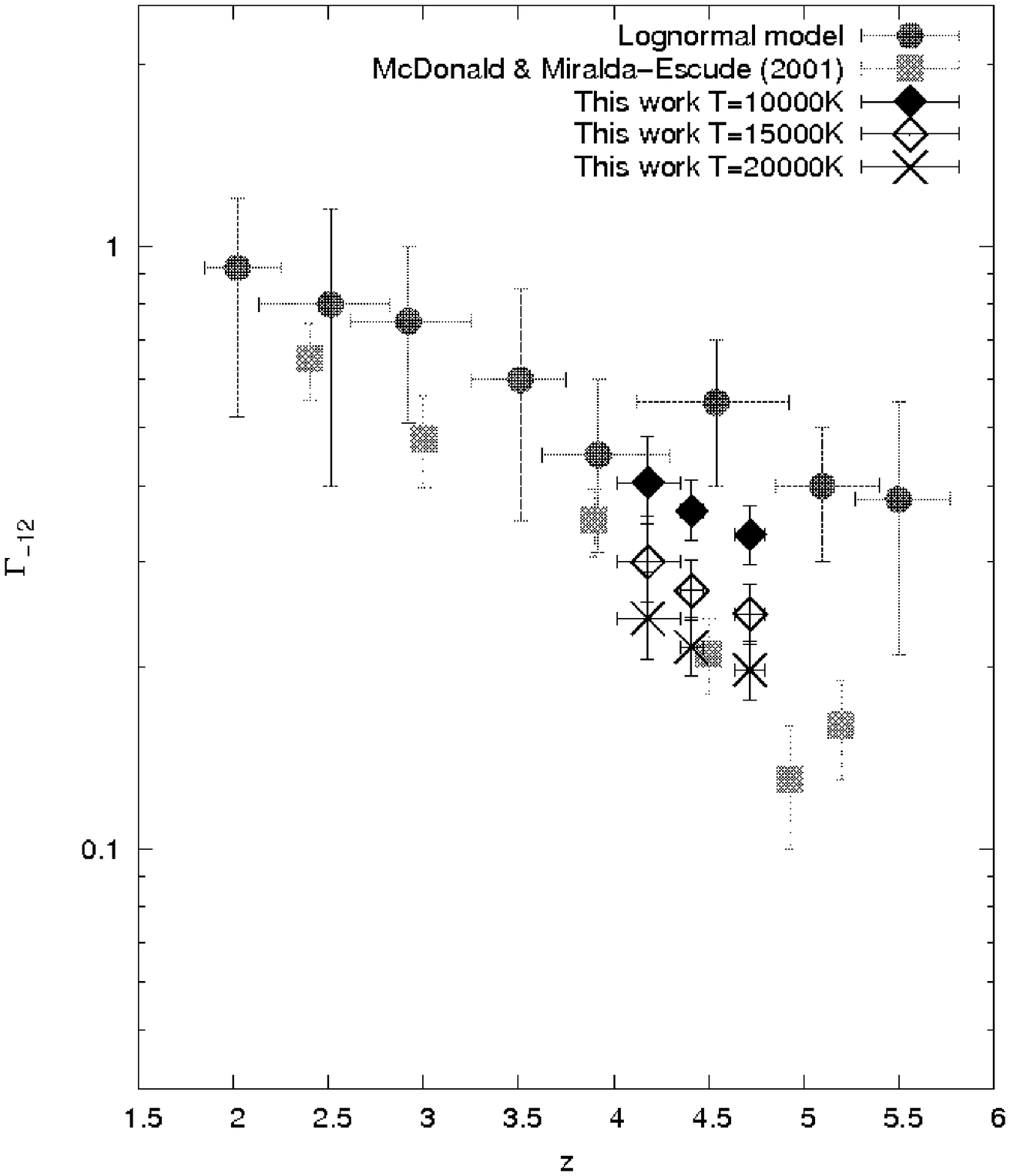}
\caption{Redshift evolution of the photoinization rate, $\Gamma_{-12}$,
derived from our data and using three different temperatures for the
IGM, $T_4$~=~1.0, 1.5 and 2, in units of $10^4$~K (filled diamond, open
diamond and crosses, respectively). Filled squares show the results by McDonald \& 
Miralda-Escud\'e (2001) and filled circles 
the results from Becker et al. (2006). $\Omega_{\rm b} h^2=  0.019$, 
$\Omega_{\rm m} = 0.3$ are assumed here.}\label{gamma}
\end{center}
\end{figure}

The normalized ionization rate defined in Eq.~\ref{geq} is derived from
the mean normalized transmitted flux by inverting Eq.~\ref{gammabg}.
Results are plotted in Fig.~\ref{g}. To decrease the errors, we have used only
three bins at $z>4$. It can be seen that our measurements
are consistent with those by McDonald et al. (2000) at lower redshift.
Fitting both McDonald's and our results together, we find that the
redshift evolution of $g$ is described as $g$~=~1.11$\times$[(1+z)/6]$^{-1.63}$. Note
that our fit is not consistent with the results by Songaila (2004)
who find  $g$~=~0.74$\times$[(1+z)/6]$^{-4.1}$ for $z>4$.
The corresponding photoionization rate, $\Gamma_{-12}$, in units of
$10^{-12}$~s$^{-1}$, is given in Fig.~\ref{gamma}, assuming $\Omega_{\rm b} h^2=  0.019$, 
$\Omega_{\rm m} = 0.3$ and three different values of gas temperature 
$T_4$~=~1, 1.5 and 2, in units of $10^4$~K. These results are consistent
with the measurements by McDonald \& Mirada-Escud\'e (2001) for a mean temperature
of $T_4 \sim$1.5 to 2 which is the mean temperature expected in the IGM at these
redshifts. However, recent determination of this quantity (Becker et al. 2006)
using a lognormal distribution for the optical depth distribution
indicate that $\Gamma_{-12}$ could be higher by about a factor of two
to three at these redshifts. We therefore have used in the following 
a mean temperature of $T_4 = $~1. 

\section{The L\lowercase{y}$\alpha$ optical depth statistics in the IGM}

To study the influence of the additional ionizing flux from
the quasar on the Ly$\alpha$ optical depth, we have
first to derive the evolution of the optical depth with redshift 
in the IGM at large distances from the quasar.
The observed HI opacity is simply:
\begin{equation}\label{opticaldepth}
\tau_{\rm HI}(\lambda) =
-{\rm ln}(\frac{F_{\lambda_{\rm obs}}}{F_{\lambda_{\rm cont}}}),
\end{equation}
where $F_{\lambda_{\rm obs}}$ is the observed flux and
$F_{\lambda_{\rm cont}}$ is the flux in the continuum.
Its evolution with redshift is described as
\begin{equation}\label{opticaldepthevolution}
\tau \propto (1+z)^\alpha.
\end{equation}

\begin{figure}
\begin{center}
\includegraphics[height=8.5cm,width=8.5cm]{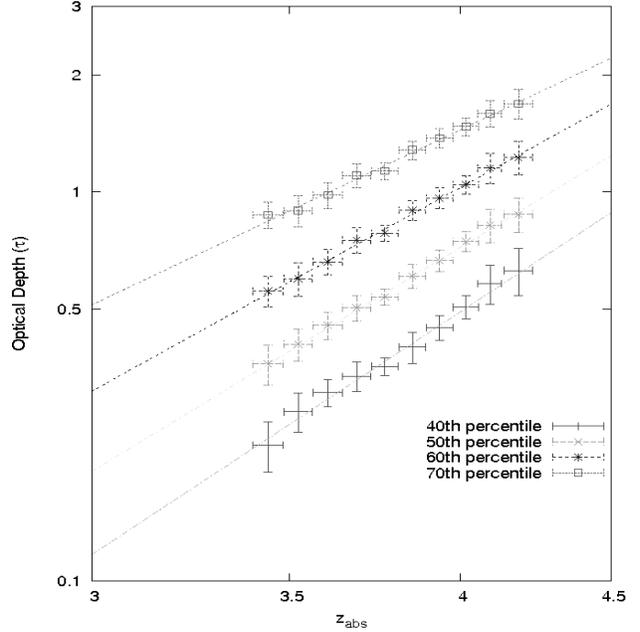}
\caption{Redshift evolution of different percentiles derived from the
CPDF of the pixel optical depth. }\label{optdepth}
\end{center}
\end{figure}

For each spectrum of the sample we estimate the optical
depth, $\tau$, in each pixel between 
$\tau_{\rm min}=-{\rm ln}(1-3\sigma)$ and $\tau_{\rm max}=-{\rm ln}(3\sigma)$, where
$\sigma(\lambda)$ is the rms of the noise measured in the spectrum (see
column 5 of Table \ref{logbook}).
We then construct the cumulative probability
distribution of $\tau$ (CPDF) in redshift windows of approximately
$\triangle z = 0.04$ corresponding to $50$~\AA~ in the observed frame
and estimate the associated percentiles.
Because low and high values of the optical depth are lost
either in the noise or because of saturation, we can use only the 
intermediate values of the percentiles.
The evolutions with redshift of the 40, 50, 60 and 70~\% percentiles
are given in Fig.~\ref{optdepth}. 

The redshift evolution index in Eq.~\ref{opticaldepthevolution}
can be derived from the redshift evolution of the percentiles of the 
pixel optical depth CPDF (see Rollinde et al. 2005).
The values obtained for $\alpha$ are ($5.8\pm 0.5$) for the 
40~\% percentile, ($5.1\pm 0.2$) for the 
50~\% percentile,($4.3\pm 0.2$) for the 60~\% percentile and ($3.8\pm 0.4$) 
for the 70~\% percentile. We use throughout the rest of the paper $\alpha$=4.8.

\section{The proximity effect from optical depth statistics }

\begin{figure}
\begin{center}
\includegraphics[height=8.5cm,width=8.5cm]{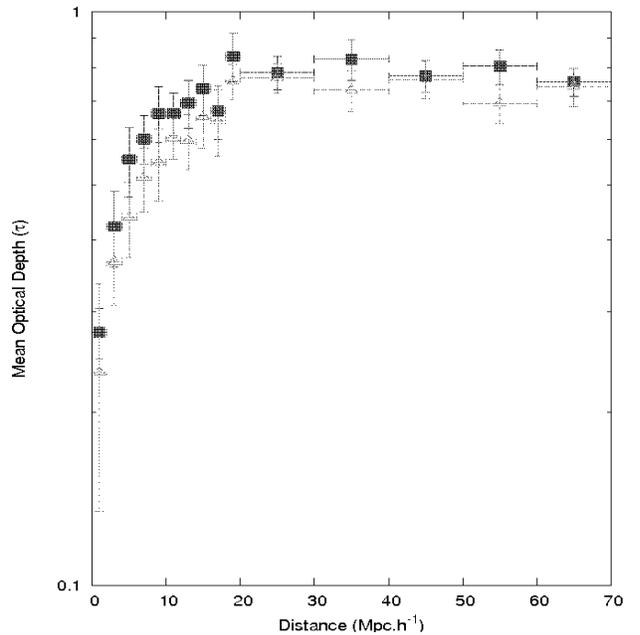}
\caption{The median optical depth corrected for redshift evolution
($\tau \propto (1+z)^{\alpha}$ with $\alpha$~=~4.8) is given versus the 
distance to the quasar, derived using QSO emission redshifts from column 3 of Table 1 (filled squares). Open triangles correspond to the case where these redshifts are increased by an amount
taken randomly between 0 and 1500~km~s$^{-1}$.}
\label{proeff}
\end{center}
\end{figure}

The evolution with redshift of the optical depth in the IGM is  
inverted in the vicinity of the quasar due to the additional
ionizing flux from the quasar. To disentangle the two 
effects we can correct the observed optical depth for the
redshift evolution in the absence of the quasar by replacing
$\tau_{\rm i}$ at redshift $z_{\rm i}$ by 
$\tau_{\rm i}$$\times$$[\frac{(1+z_{\rm ref})}{(1+z_{\rm i})}]^{\alpha}$,
where $z_{\rm ref}$ is a fixed redshift taken as reference.
We have corrected the observed optical depth for each pixel using the factor above
with $z_{\rm ref}=4.5$ and the value of $\alpha$ obtained from
the redshift evolution of the IGM CPDF, $\alpha$~=~4.8. The results are shown in Fig~\ref{proeff} 
(filled squares). The observed corrected median optical depth is given versus
the distance to the quasar computed using the cosmological 
parameters given in Section~2 and assuming emission redshifts
from column 3 of Table~\ref{logbook}. The proximity effect is apparent as a decrease of the
optical depth within a distance to the quasar smaller than $\sim$15-20~$h^{-1}$Mpc.

\subsection{QSO systemic redshifts}

The above effect depends on the accurate determination of 
the QSO systemic redshift. Gaskell (1982) has shown that the redshifts derived from different
quasar emission lines often do not agree with each other within
typical measurement errors. The high-ionization broad emission lines (HILs;
e.g., Ly$\alpha$ $\lambda1215.67$, CIV $\lambda1549$,
CIII $\lambda1909$ and NV $\lambda1240$) are found
systematically blueshifted with respect to the low-ionization
broad emission lines (LILs; e.g, OI $\lambda1305$,
MgII $\lambda2798$, and the permitted Balmer series) and
principally from the forbidden narrow emission lines
(e.g., NeV $\lambda3426$, OII $\lambda3727$,
NeIII $\lambda3870$, OIII $\lambda5007$). The narrow
emission lines arise from gas located in the galaxy host and 
their redshift should be more representative of the center-of-mass
redshift (see e.g., Tytler \& Fan 1992, Baker et al. 1994).

Unfortunately, for our sample of quasars with emission redshifts 
in the range $4 \le z_{\rm em} \le 4.5$, the forbidden narrow emission lines 
cannot  be detected from the ground because they are redshifted
in near-infrared spectral windows that are absorbed by the terrestrial 
atmosphere. We have derived the emission redshifts using 
two approachs. In the first approach, the final redshift (see column~2 of Table \ref{logbook}) 
is the mean of two estimates: one obtained by fitting a Gaussian to the CIV 
emission line and the other by measuring the peak of the Ly$\alpha$ emission line 
to avoid the various absorption features shortwards of the line. 
In the second approach, the emission redshift (see column~3 of Table \ref{logbook}) 
was obtained using the IRAF task 
rvidlines. Initially we identify a prominent spectral feature (usually the Ly$\alpha$
emission line) to which a gaussian function is fitted. Based on the central wavelength 
of this line and an input list of known spectral features, other features at 
a consistent redshift are identified and fitted. The final emission redshift is a 
weighted average value based on the gaussian fits.
The median optical depth of the IGM versus the distance to the quasar
computed using emission redshifts obtained by this second approach is
shown as squares in Fig.~\ref{proeff}. 

As we know that HILs are systematically blueshifted, with respect to
LILs and narrow emission lines, from $500$ to $1500$~km~s$^{-1}$
(Tytler \& Fan 1992) we have increased the emission redshift 
(obtained by fitting the emission lines) by a random amount and calculated
the optical depth versus the distance to the quasar as follows.
For each
realization, we increase the redshift of each of the 45 quasars by an amount
taken randomly between 0 and 1500~km~s$^{-1}$. We calculate the
distance of each pixel of the lines of sight to the corresponding 
quasar and compute the median optical depth for each value of $r$ by
averaging over all quasars. We then average the optical depths
over a hundred realizations. Errors are taken as the mean rms
obtained over the realizations. 
Results are shown in Fig.~\ref{proeff} (triangles). The proximity effect
is more pronounced in that case, as expected. This approach will be used 
in the next Sections whenever we will use QSO emission redshifts.

\subsection{The QSO ionization rate} 

The strength of the proximity effect depends on the ratio of the
ionization rates from the QSO emission and the UV-background. The QSO ionization 
rate can be determined directly from its luminosity.
The HI ionization rate due to a source of UV photons is formally
given by the equation:
\begin{equation}\label{LambdaQSO}
\Gamma_{\rm QSO}=\int_{\nu_0}^\infty [4\pi J_{\rm QSO}(\nu)\times
\frac{\sigma_{\rm HI}(\nu)}{h \nu}]. d\nu ({\rm s}^{-1})
\end{equation}
where, $\nu_0$ is the frequency of the Lyman limit,
$\sigma_{\rm HI}(\nu)=6.3\times10^{-18} (\frac{\nu_0}{\nu})^3$~cm$^2$ is the
HI photo-ionization cross-section, 
$J_{\rm QSO}(\nu) = J_{\rm QSO}(\nu_0)\times (\frac{\nu}{\nu_0})^{-\phi}$, if we assume
that the ionizing spectrum is a power law of index $\phi$ and
where, $J_{\rm QSO}(\nu_0)$ is defined as:
\begin{equation}
4\pi\,J_{\rm QSO}(\nu_0)= \frac{L_{\rm QSO}}{4\pi r^2}
\end{equation}
with $L_{\rm QSO}$ the monochromatic luminosity of the quasar
at the Lyman limit. These luminosities are computed 
extrapolating the flux in the continuum at $\lambda_{\rm obs}$~$\sim$~6000~\AA~ 
using a power-law of index $\phi = -0.6$ (Francis et al. 1993).  
We checked that within a reasonable range of $\phi = -0.5$ to $ -0.7$ 
(e.g., Cristiani \& Vio 1990), our main result 
(i.e. the  density structure around quasars) is not affected by this choice. 
Therefore,
\begin{equation}
\Gamma_{\rm QSO}=\frac{12.6\times 10^{-12}}{3+\phi}J_{\rm QSO}(\nu_0)\,10^{21} ({\rm s}^{-1})
\end{equation}
thus,
\begin{equation}
\Gamma_{\rm QSO}^{12}\,=\,
\frac{12.6}{3+\phi}\frac{L_{\rm QSO}/4\pi}{4\pi\,r^2}\,10^{21} (10^{-12}{\rm s}^{-1})
\end{equation}
$r$ is the luminosity distance from the quasar of
emission redshift $z_{\rm em}$ to the cloud at redshift $z$
and is calculated using equations given by Liske (2003) for a flat cosmological model.

\subsection{Ionizing rate ratio}
\begin{figure}
\begin{center}
\includegraphics[height=8.5cm,width=8.5cm]{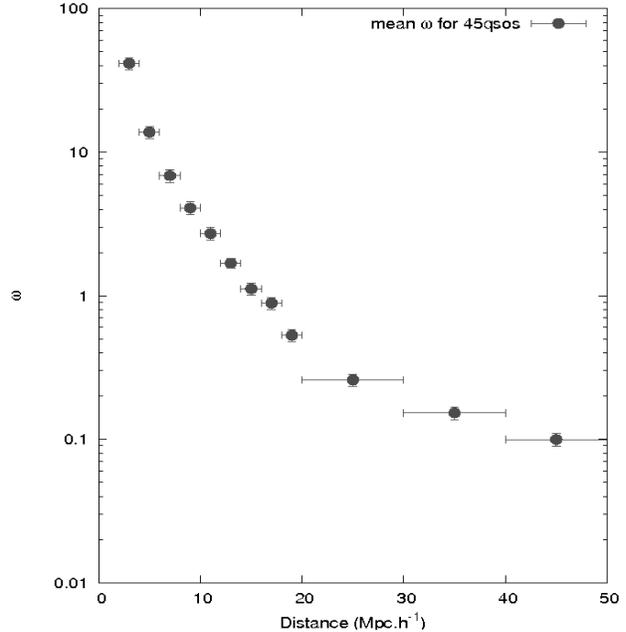}
\caption{The ratio $\omega= \frac{\Gamma_{\rm QSO}}{\Gamma_{\rm bck}}$ 
of the ionizing rate from the QSO to
that from the UV background versus the cosmological distance
between the quasar and the cloud in the IGM. $\Gamma_{\rm bck}$ is taken from Fig. \ref{gamma}.} \label{wfor45QSOs}
\end{center}
\end{figure}

We define $\omega$ as the ratio of the ionizing rate due to the quasar
emission to that due to the UV background, 
$\omega=\frac{\Gamma_{\rm QSO}(r,z)}{\Gamma_{\rm bck}(z)}$.
The later has been derived in Section~4 (see Fig.~5).
The enhanced ionizing flux in the vicinity of the QSO induces a
decrease in the optical depth of the IGM observed
in the absence of the QSO by a factor ($1+\omega$). 
For each spectrum, we have calculated $\omega_{\rm QSO}$ and $r$ for each pixel and then,
at a given $r$, we have derived $\omega$ for the whole sample as 
the median of the individual values $\omega_{\rm QSO}$(r) found for each of the 45 quasars. 
The result is shown in Fig.~\ref{wfor45QSOs}. 

\subsection{Overdensities around the quasar}

In the standard analysis of the proximity effect it is assumed
that the matter distribution in the IGM is not altered by the presence of the
quasar. The only difference between the gas located either close 
to the quasar or far away from it is the increased photo-ionization rate 
in the vicinity of the quasar. 
In that case, far away from the quasar, the median optical depth corrected for
redshift evolution, $\tau_{\rm median}$, should be a constant we can call $\tau_0$. 
In the vicinity of a QSO, due to the emission of ionizing photons,
$\tau_{\rm median}$ is no more a  constant (see Fig.~7) and decreases when $r$ 
decreases as:

\begin{equation}\label{opticaldepth}
\tau_{\rm median} = \tau_0 \frac{1}{(1+\omega)}
\end{equation}

where $\omega$ was defined and derived in the previous 
Section (see Fig.~8). 

\begin{figure}
\begin{center}
\includegraphics[height=8.5cm,width=8.5cm]{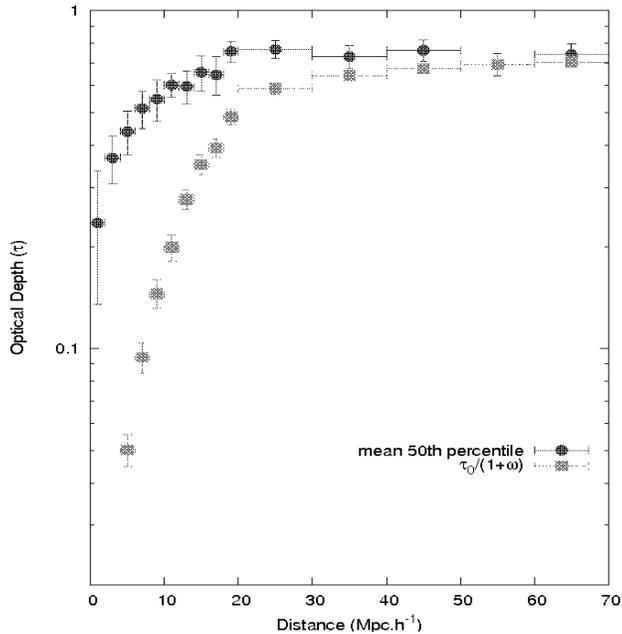}
\caption{Comparison of median optical depth's evolution observed
close to the quasars (filled circles) with that of an evolution due to the ionization field of the quasar only (Eq.~\ref{opticaldepth}; filled squares). 
The difference between the curves can be interpreted as the presence of an overdensity of gas in the
vicinity of the quasar.
} 
\label{optdepdens}
\end{center}
\end{figure}

The observed median optical depth $\tau_{\rm median}^{\rm obs}$ is compared to 
$\tau_0 / (1+\omega)$ in Fig~\ref{optdepdens}. It is apparent
that the two curves do not agree and that the observed median optical
depth is much larger than what would be expected in the case the density field of the IGM remains unperturbed when approaching
the quasar. This means that on an average, there is an overdensity 
of gas compared to the IGM in the vicinity of the quasar.

If an overdensity, $\frac{\rho(r)}{\rho_0}$, is present around
the quasar, then the median optical depth should be:
\begin{equation}\label{opticaldepth}
\tau_{\rm median} = \tau_0 \frac{(\rho(r)/\rho_0)^{[2-0.7(\gamma -1)]}}{(1+\omega)}
\end{equation}
The $\gamma$ exponent has been shown by Schaye et al. (2000) to be in the 
range $\gamma$~=~[1-1.5], thus it is reasonable to assume $\gamma$~=~1 
for this work.

\begin{figure}
\begin{center}
\includegraphics[width=8.5cm]{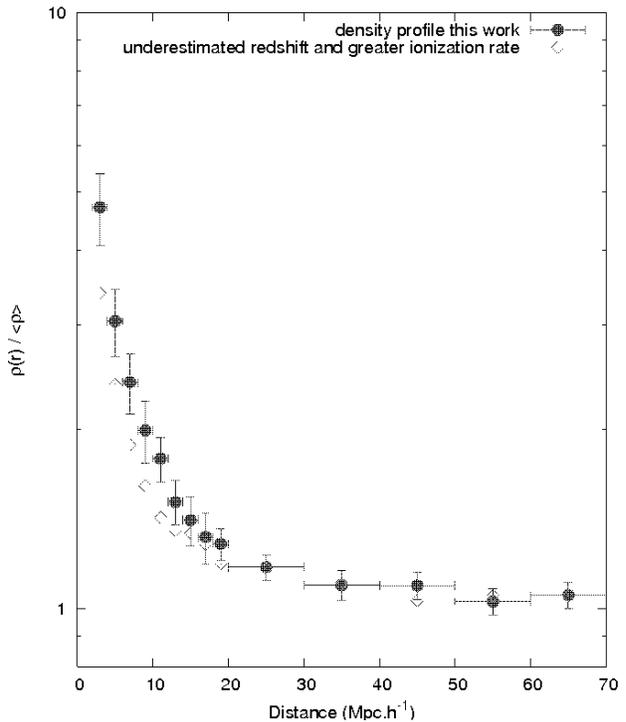}
\caption{The mean density profile around quasars with emission redshift in the
range $4 \le z_{\rm em} \le 4.5$ as derived from this work is shown as filled
circles. Open diamonds indicate the lower limit on the density profile derived
assuming that the redshift from column~3 Table~\ref{logbook} is
systematically underestimated by 1500~km~s$^{-1}$ and that the ionizing rate
in the IGM is that measured by Becker et al. (2006).}
\label{density}
\end{center}
\end{figure}

Using Eq.~\ref{opticaldepth} we can derive the actual overdensity.
It is given versus the distance to the quasar in Fig.~\ref{density} (filled circles). 
The gas density close to the quasar is significantly higher than the mean
density in the IGM at least within the first 10$h^{-1}$Mpc or so. 

\section{Discussion and Conclusions}

In this paper we have used the method presented by Rollinde et al. (2005)
to probe the density structure around quasars. In the vicinity
of the quasar and in comparison with the situation far away from it, 
the ionization factor of the gas is increased
by the emission of ionizing photons by the quasar and is decreased by the
presence of an overdensity.
If the redshift evolution of the IGM photo-ionization 
rate and that of the median optical depth in the IGM are both constrained 
by the observed evolution of the median optical depth in the 
Ly-$\alpha$ forest far from the quasar, then it is possible
to derive the density structure in the vicinity of the quasar.
We have followed McDonald \& Miralda-Escud\'e (2001) to derive
the mean ionization rate in the IGM at $z>4$. Our results are lower, by a factor
of about 1.5, than
the new results by Becker et al. (2006). We then have used the
redshift evolution of the pixel optical depth PDF to estimate 
the redshift evolution of the median optical depth of the IGM.
Using these results, we have found that quasars are surrounded
by significant overdensities on scales up to about 10~$h^{-1}$Mpc.
The overdensity is of the order of a factor of two at 10~$h^{-1}$Mpc
but is larger than five within the first megaparsec. If true, we 
can estimate that the mass surrounding 
the quasars within 1~Mpc at this redshift is of the order of $10^{14}$~M$_{\odot}$ 
corresponding to the mass of a big cluster of galaxies.
\par\noindent
An overdensity can be artificially derived from the above analysis
if (i) the photo-ionization rate is underestimated and/or
(ii) the redshift of the quasars are systematically underestimated
for any reason. To check the robustness of our result we can try to 
obtain a lower limit on the overdensity surrounding the quasars. For this
we have arbitrarily increased the photo-ionization rate that we used 
by a factor of 1.5 so that it is equal to the value obtained by Becker et al.
(2006) at the same redshift. In addition, we have increased all quasar emission 
redshifts by 1500~km~s$^{-1}$
to account for a possible systematic underestimate of the emission redshift when
using the Ly-$\alpha$ emission line (note however that in our treatment we
already increase the emission redshift from the Ly-$\alpha$ emission line
by an amount taken randomly between 0 and 1500 km~s$^{-1}$).
The result is shown in Fig.~\ref{density} (open diamonds). It can be seen 
that the overdensity is still present although smaller as expected.
\par
\begin{figure}
\begin{center}
\includegraphics[width=8cm,width=8.5cm]{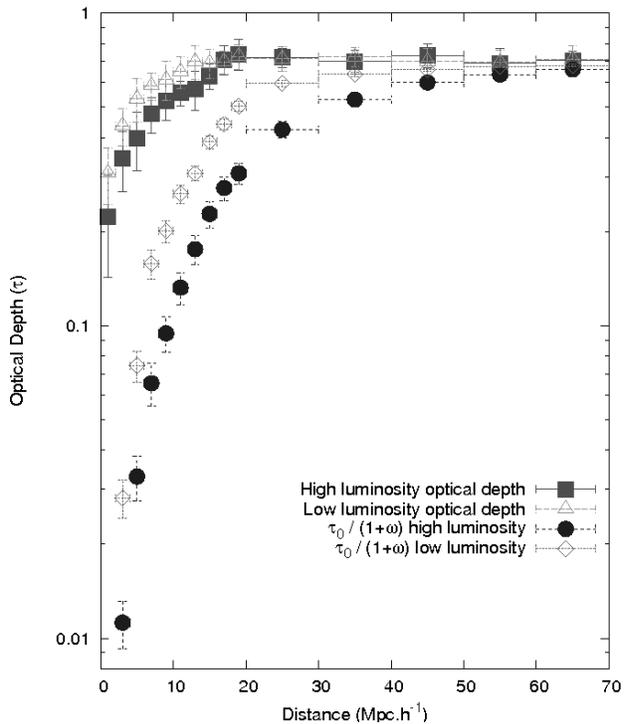}
\caption{The median optical depth in versus 
the distance from the cloud to the quasar for two subsets of our
QSO sample: one including the 15 QSOs of highest luminosities 
(filled squares) and the other 
including the 17 QSOs of lowest luminosities (open triangles). 
It is apparent that the 
proximity effect is correlated with luminosity as expected.
The curves giving the factor (1~+$\omega$) are drawn for each
subset for illustration (filled circles and open diamonds respectively.}\label{proxsubset}
\end{center}
\end{figure}

\begin{figure}
\begin{center}
\includegraphics[width=8cm,width=8.5cm]{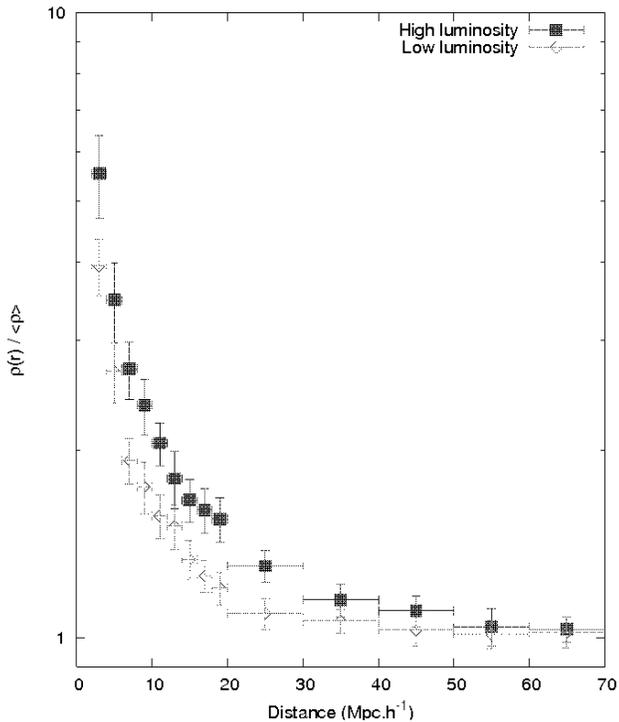}
\caption{The density profile around quasars versus 
the distance from the cloud to the quasar for two subsets of our
QSO sample: one including the 15 QSOs of highest luminosities 
(filled squares) and the other 
including the 17 QSOs of lowest luminosities (open diamonds). 
}\label{rhosubsets}
\end{center}
\end{figure}

\noindent
If our interpretation of the proximity effect is correct,
we should expect a correlation between the strength of
the observed proximity effect and the intrinsic luminosity of the QSOs. To
test this effect, we have considered two subsets of our quasar sample,
one containing the 15 QSOs with highest intrinsic luminosities 
($6.14\times10^{+31} \le L \le 1.22\times10^{+32}$~$h^{-2}$erg~s$^{-1}$Hz$^{-1}$) 
and the other containing the 17 QSOs with lowest intrinsic luminosities 
($1.85\times10^{+31} \le L \le 3.74\times10^{+31}$~$h^{-2}$erg~s$^{-1}$Hz$^{-1}$). 
We have avoided quasars with intermediate luminosities because they 
may dilute the possible result.
In Fig.~\ref{proxsubset}, we plot for both subsets the median optical
depth versus the distance to the quasar. It is apparent that the proximity effect 
is less pronounced for the low-luminosity sample compared to the high-luminosity 
sample as expected. The Proximity effect is correlated with QSO 
luminosities. For illustration, we plot also
$\tau_{0}/(1+\omega)$ for both subsets on the figure. The corresponding
overdensities around the quasars in the two subsets
are shown in Fig.~\ref{rhosubsets}. The overdensity is
correlated with luminosity. Brighter quasars show higher overdensities.
Some caution should be applied when interpreting this result however as
we have used a quite simple model to correct the observed proximity effect.
%
%

At redshift $z\sim 2$, Rollinde et al. (2005) claimed tentative
detection of overdensities of about a factor of two on scales
$\sim$5$h^{-1}$~Mpc. The result was only marginal because the statistics
was small.
With a sample of 45 quasars at $z > 4$, we have demonstrated
that overdensities exist. This result strongly supports the idea that
quasars at high redshift are located in regions of high overdensities
probably flagging the places where massive clusters of galaxies 
will form. 

It has been known for a long time that quasars are associated with
enhancements in the distribution of galaxies (Bahcall, Schmidt \&
Gunn 1961) and that at low redshift ($z<0.4$) they are associated
with moderate groups of galaxies (e.g., Fisher et al. 1996).
However, little is known at high redshift. 
Our method yields a mean one-dimensional profile of the
density distribution on large scales when other methods study the correlation of
the quasar with surrounding objects (see Kauffmann \& Haehnelt 2002). 
This result adds to the growing
evidence that high-z QSOs seem to reside in dense and probably highly
biased regions (Djorgovski 1999, Djorgovski et al. 1999).
Observational efforts should be done to obtain deep images 
of the fields around bright high-$z$ quasars
to search for the presence of any concentration of objects 
around the quasar.

\vskip 0.5cm
\noindent{\sl  Acknowledgements:} {RG is supported by a grant from the 
brazilian gouvernment (CAPES/MEC). AA is supported by a PhD grant from the 
Ministry of Science, Research \&Technology of Iran. PPJ and RS gratefully 
acknowledge support from the 
Indo-French Centre for the Promotion of Advanced Research (Centre Franco-Indien 
pour la Promotion de la Recherche Avanc\'ee) under contract No. 3004-3.
SGD is supported by the NSF grant AST-0407448,
and the Ajax Foundation.  Cataloguing of DPOSS and discovery of PSS QSOs
was supported by the Norris Foundation and other private donors.
We thank E. Thi\'ebaut, and D. Munro for freely distributing his yorick 
programming language (available at ftp://ftp-icf.llnl.gov:/pub/Yorick), which 
we used to implement our analysis. 
The authors wish to recognize and acknowledge the very significant cutural
role and reverence that the summit of Mauna Kea has always had within the
indigenous Hawaian community. We are most fortunate to have the opportunity
to conduct observations from this mountain. We acknowledge the Keck support
staff for their efforts in performing these observations.
}
\vskip 0.5cm

%

\end{document}